\documentclass[aps,prd,superscriptaddress,showpacs]{revtex4}
\usepackage{graphicx}
\usepackage{dcolumn}
\usepackage{bm}
\usepackage{amsmath}
\usepackage{float}
\usepackage[latin1]{inputenc} 

\newcommand{\be}{\begin{equation}}
\newcommand{\ee}{\end{equation}}
\newcommand{\bea}{\begin{eqnarray}}
\newcommand{\eea}{\end{eqnarray}}

\newcommand{\dl}{\delta}

\newcommand{\ep}{\epsilon}
\newcommand{\ber}{\begin{eqnarray}}
\newcommand{\eer}{\end{eqnarray}}

\newcommand{\p}{\partial}

\newcommand{\bra}[1]{\left(#1\right)}
\newcommand{\bras}[1]{\left[#1\right]}
\newcommand{\brac}[1]{\left\{#1\right\}}
\newcommand{\reff}[1]{(\ref{#1})}
\newcommand{\fat}[1]{\textbf{#1}}

\begin{document}

\title{The influence of strong field vacuum polarization on
gravitational-electromagnetic wave interaction.}
\begin{abstract}
The interaction between gravitational and electromagnetic waves in the
presence of a static magnetic field is studied. The field strength of the static
field is allowed to surpass the Schwinger critical field, such that the
quantum electrodynamical (QED) effects of vacuum polarization and
magnetization are significant. Equations governing the interaction are
derived and analyzed. It turns out that the energy conversion from
gravitational to electromagnetic waves can be significantly altered due to
the QED effects. The consequences of our results are discussed.
\end{abstract}

\author{M. Forsberg}
\affiliation{Department of Physics, Ume{\aa } University, SE--901 87 Ume{\aa }, Sweden}
\author{D. Papadopoulos}
\affiliation{Department of Physics, Section of Astrophysics, Astronomy and Mechanics,
54124 Thessaloniki,Greece}
\author{G. Brodin}
\affiliation{Department of Physics, Ume{\aa } University, SE--901 87 Ume{\aa }, Sweden}

\date{\today}
\maketitle



\section{Introduction}

As studied by many authors \cite%
{Moortgat2003,Isliker,bmd1,ignatev,bmd2,brodinmarklund,Balakin2003,BMS2001,Papadouplous2001,kallberg2004,servinbrodin,Servin2000,Papadoupolus2002,MDB2000,Mosquera2002,Moortgat2006}
there exist numerous mechanisms for the conversion between gravitational
waves (GWs) and electromagnetic (EM) waves. In particular, the propagation
of GWs across an external static magnetic field gives rise to a linear
coupling to the electromagnetic field (see e.g. Ref. \cite%
{Moortgat2003,Isliker,bmd1}), which may lead to the GW excitation of ordinary EM waves in vacuum, or of magnetohydrodynamic (MHD)
waves in a plasma \cite{Papadouplous2001,Isliker,kallberg2004}. Many
nonlinear coupling mechanisms are also possible \cite%
{ignatev,Papadouplous2001,bmd2,brodinmarklund,Balakin2003,BMS2001}.
Conversion of energy from gravitational to electromagentic degrees of
freedom has been pointed out as a means to indirect detection of
gravitational waves by several authors (see e.g. Refs. \cite{ignatev,
servinbrodin, Isliker}), since the latter is so much easier to detect. For
astrophysical application (see e.g. Refs. \cite%
{BMS2001,Mosquera2002,Moortgat2003,Moortgat2006,Isliker,ignatev}), naturally
this requires well developed theories to recognize the signature of the
gravitational origin. Furthermore, there must be a sufficient amount of
energy conversion taking place. Specifically, considering the coupling due
to a static magnetic field, it has been noted that more energy can be
converted from gravitational to electromagnetic degrees of freedom if the
interaction region is larger, and if the magnitude $B_{0}$ of the static
magnetic field is larger \cite{Isliker}. In the case that the interaction
region is magnetized vacuum, with a size smaller than the background
curvature radius, it has been found that the energy converted is linear in
the background field energy density \cite{bmd1,Isliker}. This result,
however, does not account for quantum electrodynamic (QED) vacuum
polarization effects \cite{Marklund-review,Brodin-pla,Lundin-2009,Valluri}, which
become significant when $B_{0}$ approaches the value $E_{\mathrm{cr}}/c$, where $%
E_{\mathrm{cr}}\equiv m_{e}^{2}c^{3}/\hbar e\simeq 10^{18}\mathrm{V/m}$ is
the Schwinger critical field, $m_{e}$ is the electron mass, $e$ is the
elementary charge, $c$ is the speed of light in vacuum, and $h=2\pi \hbar $
is the Planck constant.

In the present paper we will investigate the QED influence on
gravitational-electromagnetic interaction in a static field $B_{0}$ that may
be stronger than the characteristic QED scale $E_{\mathrm{cr}}/c$. It should
be noted that such intense field do occur in nature, specifically close to
magnetars where close to the surface the magnetic field strength may reach $%
10^{10}-10^{11}\mathrm{T}$ \cite{Magnetar}. Starting from Einstein's
equations, together with the Heisenberg-Euler Lagrangian to describe vacuum
polarization and magnetization in the electromagnetic theory, the basic
equations for small amplitude wave propagation on a background with a strong
static magnetic field $B_{0}$ is derived. In order to simplify the
calculation, the size of the interaction region is assumed to be much
smaller than the background curvature. It is found that the vacuum
polarization effects lead to a saturation, such that the energy conversion
(almost) stops to grow with $B_{0}$ beyond a certain value $B_{\mathrm{sat}}$%
. This value depend on the length $L$ of the interaction region. For a large
$L$, the saturation value is much smaller than the QED scale, i.e. $B_{%
\mathrm{sat}}\ll E_{\mathrm{cr}}/c$ (in which case the weak field
QED corrections \cite{Marklund-review} of the Heisenberg-Euler theory would have
sufficed), but for shorter interaction regions we may have $E_{\mathrm{cr}%
}/c\ll B_{\mathrm{sat}}$ in which case the full theory is required. The relevance of our model calculation to
astrophysical problems is discussed at the end of the paper.

\section{Basic Equations}
The Lagrangian for soft photon (i.e. photon energy much smaller than electron rest mass energy) light propagation,  taking one loop corrections into account, is given by \cite{Lundin-2009,Valluri,Schwinger}
%
\be
	\mathcal{L} = -\frac{1}{\mu_0} \mathcal{F}
	- \frac{\alpha}{2\pi \mu_0 e^2} \int_{0}^{i\infty }
	\frac{ds}{s^{3}}e^{-e E_{cr} s / c }
	\times \bras{ (es)^{2}ab\coth {(eas)}\cot {(ebs)}
	- \frac{(es)^{2}}{3}(a^{2}-b^{2})-1} - A_{\alpha }j^{\alpha } \ , \label{Lagrangian}
\ee
where $a=[\sqrt{(\mathcal{F}^{2}+\mathcal{G}^{2})}+\mathcal{F}]^{1/2}$,
$b=[\sqrt{(	\mathcal{F}^{2}+\mathcal{G}^{2})}-\mathcal{F}]^{1/2}$
%
, $\mathcal{F}=(1/4)F_{\alpha \beta }F^{\alpha \beta}$
, $\mathcal{G}=(1/4)F_{\alpha \beta }\hat{F}^{\alpha \beta }$
, $F^{\alpha \beta }$ is the electromagnetic field tensor, $\hat{F}^{\alpha \beta }=\epsilon ^{\alpha \beta \mu \nu }\frac{F_{\mu \nu }}{2}$, $\epsilon ^{\alpha \beta \mu \nu }$ the totally antisymmetric tensor, $A_{\alpha }$ the four-potential, $j^{\alpha }$ the four-current and $\alpha$ the fine structure constant.
%
The Euler-Lagrange equations of motion for the Lagrangian \reff{Lagrangian}
becomes
\begin{equation}
	\gamma_{\mathcal{F}}F_{;\mu }^{\mu \nu}+\gamma_{\mathcal{G}}\hat{F}_{;\mu}^{\mu \nu }
	+ \frac{1}{2}[\gamma _{\mathcal{F}\mathcal{F}}F^{\mu \nu}F_{\alpha \beta }
	+ \gamma_{\mathcal{G}\mathcal{G}} \hat{F}^{\mu \nu }\hat{F}_{\alpha \beta }]
	F_{,\mu }^{\alpha \beta } + \gamma_{\mathcal{F}\mathcal{G}}
	\left[ F^{\mu \nu }\hat{F}_{\alpha \beta }
	+ \hat{F}^{\mu \nu }F_{\alpha \beta}\right] F_{,\mu }^{\alpha \beta}= - j^{\nu } \ ,
	\label{QED-eqs-of-motion}
\end{equation}%
where we have applied the Eq. (2) of Ref. \cite{Lundin-2009} to a curved background, and introduced the
quantities
\begin{eqnarray}
	&&\gamma _{\mathcal{F}}=\frac{\partial \mathcal{L}}{\partial \mathcal{F}}%
	,~~\gamma _{\mathcal{G}}=\frac{\partial \mathcal{L}}{\partial \mathcal{G}},
	\notag \\
	&&\gamma _{\mathcal{F}\mathcal{F}}=\frac{\partial ^{2}\mathcal{L}}{\partial
	\mathcal{F}^{2}},~~\gamma _{\mathcal{G}\mathcal{G}}=\frac{\partial ^{2}%
	\mathcal{L}}{\partial \mathcal{G}^{2}},  \notag \\
	&&\gamma _{\mathcal{F}\mathcal{G}}=\frac{\partial ^{2}\mathcal{L}}{\partial
	\mathcal{F}\partial \mathcal{G}} \ . \label{QED-parameters}
\end{eqnarray}%
The physics of strong field vacuum polarization and vacuum magnetization is
thus encoded in the parameters introduced in Eq. (\ref{QED-parameters}). For
the case of interest to us, i.e. no external electric field, the scalars,$%
\gamma _{\mathcal{F}},\gamma _{\mathcal{G}},\gamma _{\mathcal{F}\mathcal{F}%
},\gamma _{\mathcal{G}\mathcal{G}}$ and $\gamma _{\mathcal{F}\mathcal{G}}$
can be computed analytically as functions of the external constant
magnetic field strength $B_{0}$. This procedure which involves the solution
of numerous integrals is described in Ref. \cite{Lundin-2009}, and the explicit expressions of the scalars can be found in Appendix A.

In the paper we will study the influence of a GW on a strong magnetic field. The metric of a linearized GW propagating in the $z$%
-direction can be written
\begin{equation}
	ds^{2} = -c^{2}dt^{2}+\left( 1+h_{+}\right) dx^{2}+\left( 1-h_{+}\right)
					dy^{2}+2h_{\times }dxdy + dz^{2}  \label{e17}
\end{equation}%
where the two independent polarizations $h_{+}$ and $h_{\times }$ depend on
the coordinates as $h_{+,\times }=h_{+,\times }(z-ct)$. Furthermore, we define an
orthonormal tetrad by
\begin{eqnarray}
	\mathbf{e}_{0} &=&\frac{1}{c}\partial _{t},  \notag \\
	\mathbf{e}_{1} &=&\left( 1-\frac{1}{2}h_{+}\right) \partial _{x}-\frac{1}{2}%
	h_{\times }\partial _{y},  \notag \\
	\mathbf{e}_{2} &=&\left( 1+\frac{1}{2}h_{+}\right) \partial _{y}-\frac{1}{2}%
	h_{\times }\partial _{x},  \notag \\
	\mathbf{e}_{3} &=&\partial _{z}\ .  \label{tetrad}
\end{eqnarray}%
In linearized theory of gravity, the relevant components of the Einstein equations read:
\begin{equation}
		(e_{0}^{2} - \partial _{z}^{2})h_{+}=\kappa (\delta T_{11} - \delta T_{22}),~~(e_{0}^{2}-\partial _{z}^{2})h_{\times }=2\kappa (\delta T_{12}) \label{gw},
\end{equation}%
where $\kappa =  8 \pi G / c^4$, and $G$ is the gravitational constant.
The energy-momentum tensor associated with the Lagrangian (\ref{Lagrangian})
is written $T_{\mu \nu }=-\gamma _{\mathcal{F}}F_{\mu }^{\alpha }F_{\alpha
\nu }+(\mathcal{G}\gamma _{\mathcal{G}}-\mathcal{L})g_{\mu \nu }$, see \cite{Gies}, and
expressions for $\delta T_{11}$, $\delta T_{22}$ and $\delta T_{12}$,
linearized around the strong magnetic field $B_{0}$, is worked out in
Appendix A. 

Next we follow the covariant approach presented in Ref. \cite{EllisElst} for splitting the EM and material fields in a 1 + 3 fashion. Suppose an observer moves with 4-velocity $u^\alpha$. This observer will measure the electric and magnetic fields $E_\alpha \equiv F_{\alpha \beta} u^\beta$ and $B_\alpha \equiv \ep_{\alpha \beta \gamma} F^{\beta \gamma} / 2$, respectively, where $F_{\alpha \beta}$ is the EM field tensor and $\ep_{\alpha \beta \gamma}$ is the volume element on hyper-surfaces orthogonal to $u^\alpha$. We also define the spatial gradient operator as $\nabla =(e_{1},e_{2},e_{3})$.
Using the $1 + 3$ split we write the Maxwell equations in the tetrad basis (\ref{tetrad}). From Eq. (\ref{QED-eqs-of-motion}) and the Faraday equation, $F_{[ij;k]}=0$, we obtain 
\begin{eqnarray}
		c\nabla \cdot \mathbf{B} &=&\frac{\rho _{B}}{\epsilon _{0}}  \label{maxwell1} \ , \\
		\nabla \cdot \mathbf{E} &=&\frac{1}{\epsilon _{0}}\left( \frac{\rho }{\gamma_{F}}+\rho _{E}\right)   \label{maxwell2} \ , \\
		e_{0}\mathbf{B}+\frac{\nabla \times \mathbf{E}}{c} &=&-\mu _{0}\mathbf{j}_{B} \label{maxwell3} \ , \\
		\frac{1}{c}e_{0}\mathbf{E}-\nabla \times \mathbf{B} &=&-\mu _{0}\left( \mathbf{j}_{Q}+\frac{\mathbf{j}}{\gamma _{F}}+\mathbf{j}_{E}\right) \ , \label{maxwell4}
\end{eqnarray}%
where $\mathbf{j}_{Q}$ is the combined vacuum polarization and vacuum
magnetization current density, which from Eq. (\ref{QED-eqs-of-motion}) can
be seen to take the form
\begin{equation}
j_{Q}^{\alpha }\equiv -\frac{1}{2\mu _{0}}\left( \frac{\gamma _{GG}}{\gamma
_{F}}\hat{F}^{kl}\hat{F}^{i\alpha }+\frac{\gamma _{FF}}{\gamma _{F}}%
F^{kl}F^{i\alpha }\right) e_{i}F_{kl},  \label{polarization_current}
\end{equation}%
and the effective (i.e. gravity induced) charge densities and current
densities are
\begin{eqnarray}
&&\rho _{E}\equiv -\epsilon _{0}\left[ \gamma _{\beta \alpha }^{\alpha
}E^{\beta }+\epsilon ^{\alpha \beta \gamma }\gamma _{\alpha \beta
}^{0}cB_{\gamma }\right] ,  \notag \\
&&\rho _{B}\equiv -\epsilon _{0}\left[ \gamma _{\beta \alpha }^{\alpha
}cB^{\beta }-\epsilon ^{\alpha \beta \gamma }\gamma _{\alpha \beta
}^{0}E_{\gamma }\right] ,  \notag \\
&&j_{E}^{\alpha }\equiv \frac{1}{\mu _{0}}\left[ -(\gamma _{0\beta }^{\alpha
}-\gamma _{\beta 0}^{\alpha })\frac{E^{\beta }}{c}+\gamma _{0\beta }^{\beta }%
\frac{E^{\alpha }}{c}-\epsilon ^{\alpha \beta \gamma }(\gamma _{0\beta
}^{0}B_{\gamma }+\gamma _{\beta \gamma }^{\delta }B_{\delta })\right] ,
\notag \\
&&j_{B}^{\alpha }\equiv \frac{1}{\mu _{0}}\left[ -(\gamma _{0\beta }^{\alpha
}-\gamma _{\beta 0}^{\alpha })B^{\beta }+\gamma _{0\beta }^{\beta }B^{\alpha
}+\epsilon ^{\alpha \beta \gamma }\left( \gamma _{0\beta }^{0}\frac{%
E_{\gamma }}{c}+\gamma _{\beta \gamma }^{\delta }\frac{E_{\delta }}{c}%
\right) \right] ,  \label{induced_currents}
\end{eqnarray}%
where the Greek indices takes values between $1$ and $3$, and the Latin
indices between $0$ and $3$. From here on we will be concerned with a
GW wave propagating across a magnetic field. Explicit expressions
of the source terms for this case is obtained by substituting the
QED-parameters from Appendix A into Eq. (\ref{polarization_current}), and
the rotation coefficients for a linearized GW presented in Appendix B into
Eq. (\ref{induced_currents}).

\section{Wave Interaction}

The most efficient interaction of a GW with a static magnetic field occurs
if the GW propagates perpendicular to the magnetic field. As has been found
by e.g. Refs. \cite{bmd1,Isliker}, the fact that the GW fulfills the same
dispersion relation as EM-waves, makes the energy conversion resonant. As a
consequence, the energy conversion from a GW to co-propagating EM-waves is
directly proportional to the background field energy density as well as the
length of the interaction region, defined as the region occupied by the
static magnetic field $B_{0}$. This conclusion holds as long as QED effects
is negligible, and the length of the interaction region is smaller than the
radius of curvature associated with the magnetic field energy density. Our
aim here is to investigate to what extent the QED effects, associated with
fields strengths approaching the Schwinger limit, modifies the energy
conversion between GW:s and EM-waves. For this purpose we will still assume
that the interaction region is smaller than the radius of curvature due to $%
B_{0}$, such that the interaction can be considered as taking place on a
Minkowski background.

As we will see, in addition to an EM-wave co-propagating with the
monochromatic GW, with metric perturbation $h_{\times ,+}=\tilde{h}_{\times
,+}\exp \left[ i(kz-\omega t)\right] $ and $\omega =kc$, a counter-propagating wave with the same frequency will also be induced. We thus make
the ansatz $\mathbf{B}=B_{0}\mathbf{e}_{1}+\delta \mathbf{B}(z)\exp \left[
-i\omega t\right] $ and $\mathbf{E}=\delta \mathbf{E}(z)\exp \left[ -i\omega
t\right] $, where $\delta \mathbf{B}$ and $\delta \mathbf{E}$ includes both
positive (along $z$) and negative propagating waves. Taking the curl of the (%
\ref{maxwell4}) and using (\ref{maxwell3}) one obtains,
\begin{equation}
-e_{0}^{2}\mathbf{B}-\nabla \times \left( \nabla \times \mathbf{B}\right)
+\mu _{0}\nabla \times \mathbf{j}_{Q}=-\mu _{0}\nabla \times \mathbf{j}%
_{E}+\mu _{0}e_{0}\mathbf{j}_{B},  \label{wave-eq-B}
\end{equation}%
to linear order, with the components of the polarization current Eq. (\ref%
{polarization_current}) given by
\begin{equation}
j_{Q}^{1}=-\frac{\gamma _{GG}}{\gamma _{F}}B_{0}^{2}\frac{1}{c\mu _{0}}%
e_{0}\delta E_{1}\ ,\ j_{Q}^{2}=-\frac{\gamma _{FF}}{\gamma _{F}}B_{0}^{2}%
\frac{1}{\mu _{0}}\partial _{z}\delta B_{1}\ ,\ j_{Q}^{3}=0.
\label{QED-current_2}
\end{equation}%
From Eq. (\ref{induced_currents}) and Eqs. \reff{ricci} the gravitational
contribution is found to be:
\begin{eqnarray}
&&\rho _{E}=\rho _{B}=0,  \notag \\
&&j_{E}^{1}=\frac{B_{0}}{2\mu _{0}}\frac{\partial {h}_{\times }}{\partial z}%
,~~j_{E}^{2}=-\frac{B_{0}}{2\mu _{0}}\frac{\partial {h}_{+}}{\partial z}%
,~~j_{E}^{3}=0,  \notag \\
&&j_{B}^{1}=-\frac{B_{0}}{2c\mu _{0}}\dot{h}_{+},~~j_{B}^{2}=-\frac{B_{0}}{%
2c\mu _{0}}\dot{h}_{\times },~~j_{B}^{3}=0.  \label{induced_currents_2}
\end{eqnarray}%
Using Eqs. (\ref{maxwell3}), (\ref{wave-eq-B}), (\ref{QED-current_2}) and (%
\ref{induced_currents_2}) we will next demonstrate that different EM wave polarizations
couple to different GW polarizations. The result is most easily expressed in terms of the
magnetic field components, and can then be written: 
\begin{eqnarray}
\left[ {k_{E}^{+}}^{2}+\partial _{z}^{2}\right] \delta B_{1} &=&{k_{E}^{+}}%
^{2}B_{0}\tilde{h}_{+}\exp \left[ ikz\right]  \notag \\
\left[ {k_{E}^{\times }}^{2}+\partial _{z}^{2}\right] \delta B_{2} &=&\frac{1%
}{2}\left( \frac{\omega ^{2}}{c^{2}}+{k_{E}^{\times }}^{2}\right) B_{0}%
\tilde{h}_{\times }\exp \left[ ikz\right] ,  \label{wave_equation_final}
\end{eqnarray}%
where ${k_{E}^{+}}^{2}=\omega ^{2}/\left( c^{2}\left( 1+B_{0}^{2}\gamma
_{FF}/\gamma _{F}\right) \right) $ and ${k_{E}^{\times }}^{2}= \omega^{2} \left( 1-B_{0}^{2}\gamma _{GG}/\gamma _{F}\right)/c^{2} $. As
can be seen, all effects of the QED-vacuum polarization and magnetization is
encoded in the effective wave-numbers ${k_{E}^{+}}$ and ${k_{E}^{\times }}$,
that approach $\omega /c$ for $cB_{0}/E_{\mathrm{cr}}\ll 1$. Note that Eq. (%
\ref{wave_equation_final}) agrees with Ref. \cite{Lundin-2009}, when the
GW-coupling terms on the right hand sides are dropped \cite{polarization-comment}.
%
%
The backreaction on the GW can be obtained by combining Eqs. \reff{gw} and \reff{EM_tensor_variation}. Whether or not this effect is important depends on the ratio of the
excited wave energy density compared to the (pseudo) wave energy density of
the GW. Roughly the scaling is as follows: For weak background magnetic
fields (i. e. negligible QED effects), the excited wave energy density is
limited by
$W_{\mathrm{em}}\sim B_{1}^{2}/\mu _{0}\sim (kL)^{2}|
\tilde{h}_{+,\times }|^{2}B_{0}^{2}/\mu _{0}$
, where $k$ is the
incident wave number and $L$ is the length of the interaction
region. As we will see in the next section, whenever QED effects are
important, the excited wave energy is reduced compared to this scaling. Thus
at most the the ratio of the excited wave energy to the GW (pseudo) wave
energy density becomes $W_{\mathrm{em}}/W_{\mathrm{GW}}=L^{2}(G/8\pi
c^{2})B_{0}^{2}/\mu _{0}$. Whenever the interaction region is smaller than
the background curvature due to the unperturbed magnetic field (as we have
assumed above), this ratio is much smaller than unity, and hence the
backreaction on the GW can be neglected. As a consequence, the approximation
of "no GW back-reaction" will be employed in the next section.

\section{A specific Example}

\begin{figure}[tbph]
\includegraphics[width={6cm}]{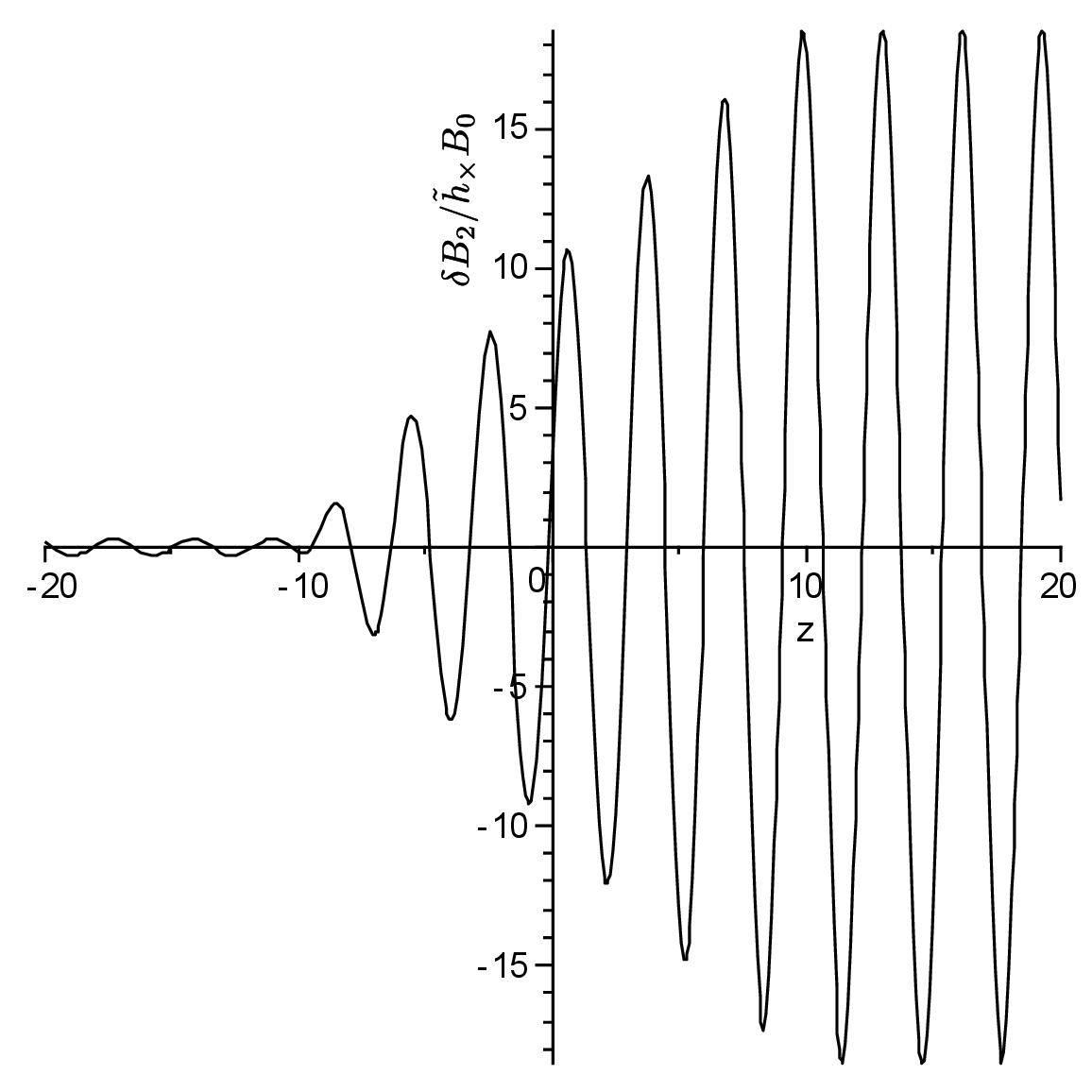}
\caption{The wave profile for $kL = 40 $ and $ c B_0 / E_{cr} = 100 $. The magnetized region lies between $z = -10$ and $ z = 10$.  }
\label{wave_profile}
\end{figure}

\begin{figure}[tbph]
\includegraphics[width={6cm}]{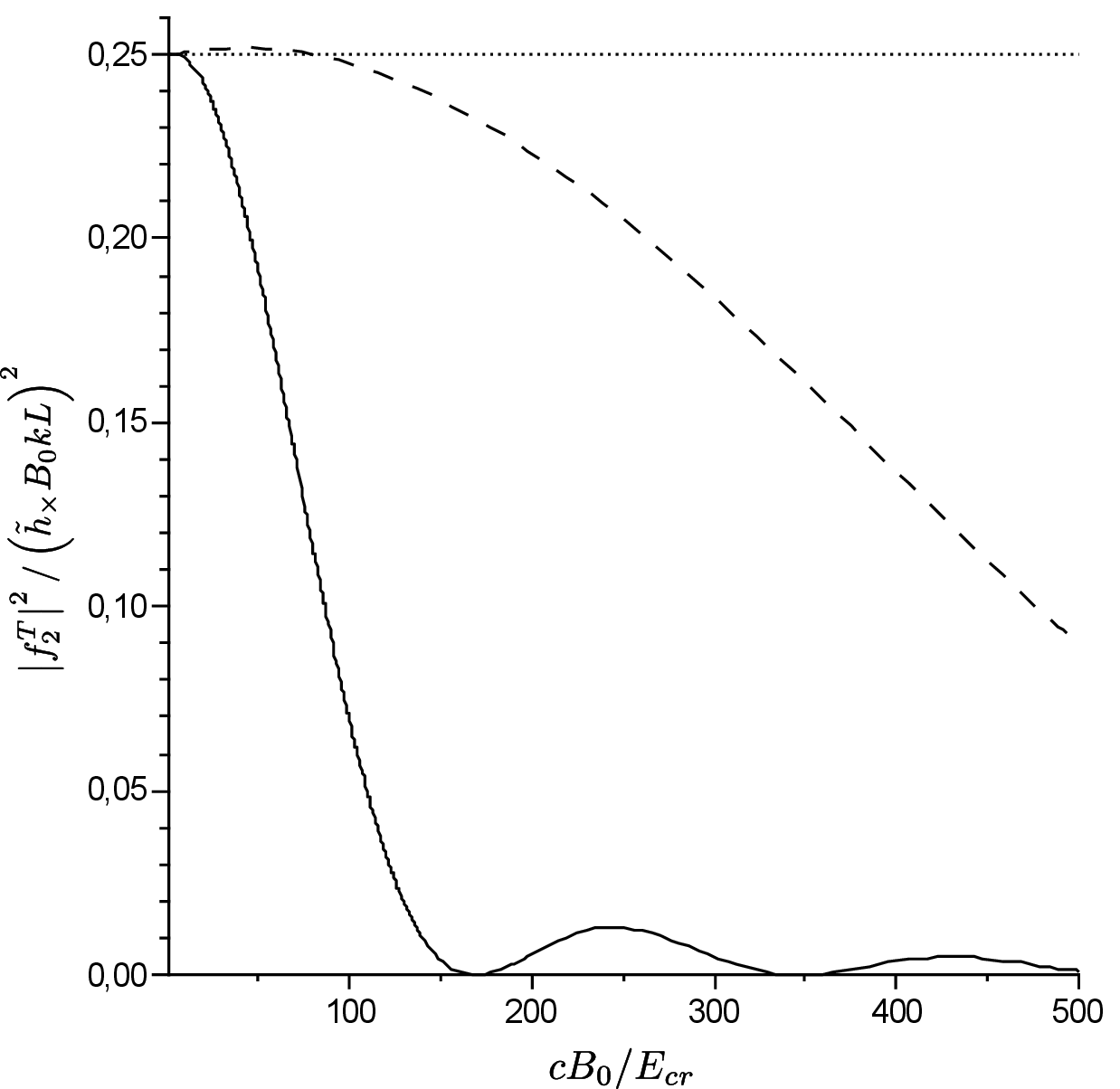}
\caption{Normalized energy density of the copropagating EM-wave, excited by a
cross-polarized GW, as a function of background magnetic field strength for $%
\protect\theta =20$ (dashed line) and $\protect\theta = 100$ (solid line) compared to the non-QED-case (dotted line).}
\label{Twavecross}
\end{figure}

\begin{figure}[tbph]
\includegraphics[width={6cm}]{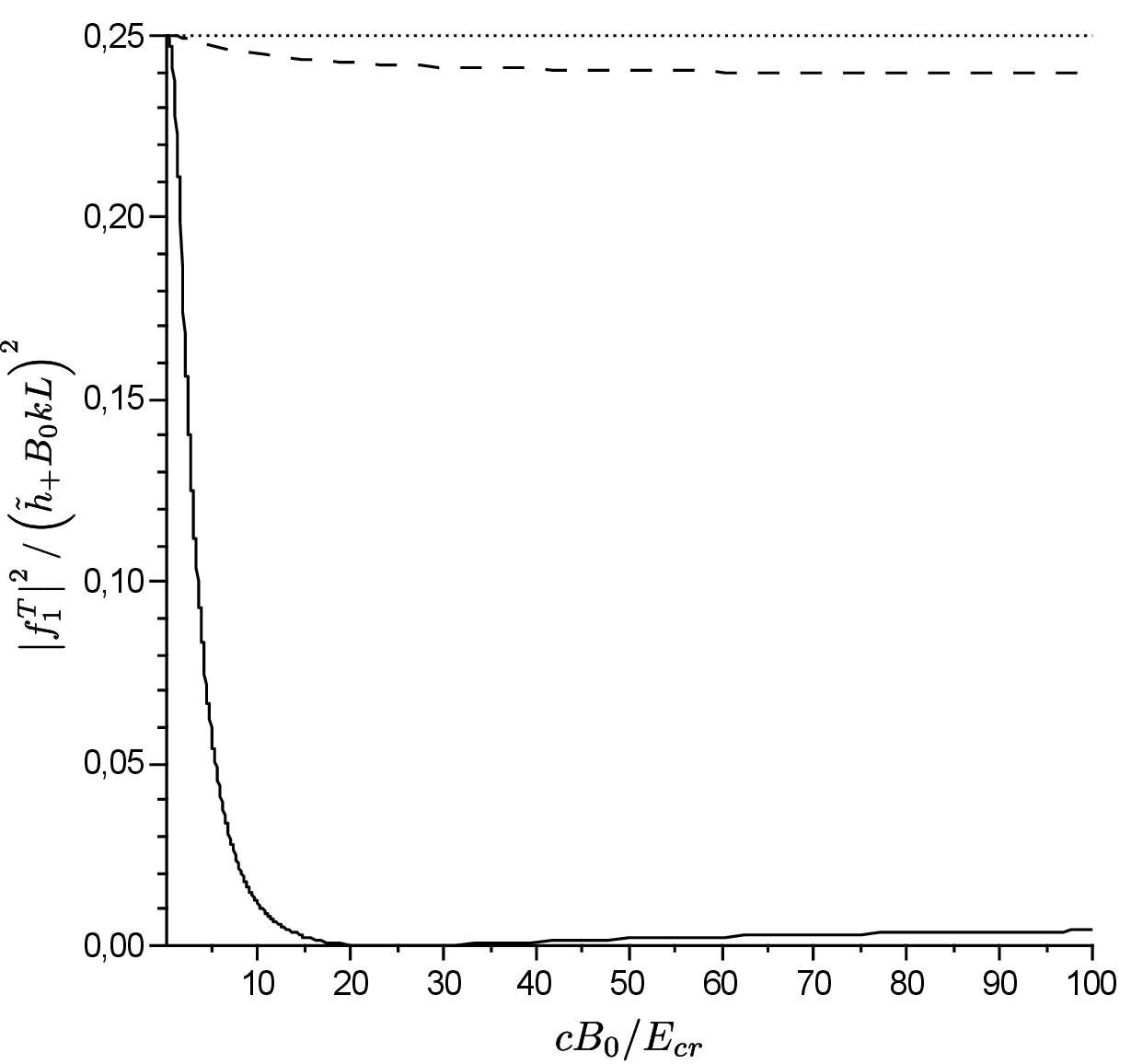}
\caption{Normalized energy density of the copropagating EM-wave, excited by a
plus-polarized GW, as a function of background magnetic field strength for $%
\protect\theta =2000$ (dashed line) and $\protect\theta =20000$ (solid line) compared to the non-QED-case (dotted line).}
\label{Twaveplus}
\end{figure}

As a specific example we will now consider a boundary value problem, where
the GW propagating in the $\mathbf{e}_{3}$-direction, enters the interaction
region, given by $-L/2<z<L/2$, which is the region where the external
magnetic field $B_{0}\mathbf{e}_{1}$ is taken to be nonzero.
The general solution to Eq. (\ref{wave_equation_final}), for the interaction
region $-L/2<z<L/2$, is
\begin{equation}
\delta B_{1,2}=T_{1,2}e^{ik_{E}^{+,\times }z}+R_{1,2}e^{-ik_{E}^{+,\times
}z}+C_{1,2}e^{ikz},  \notag
\end{equation}%
where $C_{1}={k_{E}^{+}}^{2}B_{0}\tilde{h}_{+}/({k_{E}^{+}}^{2}-k^{2})$,
$C_{2}=({k_{E}^{\times }}^{2}+k^{2})B_{0} \tilde{h}_{\times }
/ 2( { k_{E}^{\times }}^{2}-k^2 ) $, and $R_{1,2}$ and $T_{1,2}$ are constants determined by the boundary conditions. This must be matched with the EM wave solutions
with constant amplitudes outside the interaction region
\begin{eqnarray}
	\delta B_{1,2}=f_{1,2}^{R}e^{-i k z} &,z\in &\left( -\infty ,-L/2\right) ,
	\notag \\
	\delta B_{1,2}=f_{1,2}^{T}e^{i k z} &,z\in &\left( L/2,\infty \right) ,
	\notag
\end{eqnarray}%
at $z=\pm L/2$.
%
%
Furthermore, the electric fields must be matched as well. The relevant
Maxwell equations are
\begin{eqnarray}
\delta E_{2} &=&\frac{i}{\omega }\left( \frac{\omega ^{2}}{{k_{E}^{+}}^{2}}%
\partial _{z}\delta B_{1}+\frac{B_{0}}{2}\partial _{z}h_{+}\right) \ , \\
\delta E_{1} &=&-\frac{i\omega }{{k_{E}^{\times }}^{2}}\left( \partial
_{z}\delta B_{2}+\frac{B_{0}}{2}\partial _{z}h_{\times }\right) \ ,
\end{eqnarray}%
The matching of the electric field is done in the same way as that of the
magnetic field to give four equations for four quantities, for each set of
coupled polarizations. Solving these equations, the resulting amplitudes of
the "reflected" and "transmitted" (or strictly speaking counter-propagating and co-propagating)
EM-waves becomes
\begin{equation}
f_{1}^{R}=\frac{B_{0}\tilde{h}_{+}}{2}\eta _{+}e^{-i\theta }\frac{\left(
1+\eta _{+}\right) e^{i\eta _{+}\theta }+\left( 1-\eta _{+}\right) e^{-i\eta
_{+}\theta }-2e^{i\theta }}{\left( 1+\eta _{+}\right) ^{2}e^{-i\theta \eta
_{+}}-\left( \eta _{+}-1\right) ^{2}e^{i\theta \eta _{+}}},
\label{reflected_amplitude_+_z}
\end{equation}%
and%
\begin{equation}
f_{1}^{T}=\frac{B_{0}\tilde{h}_{+}}{2}\frac{\eta _{+}}{\left( 1-\eta
_{+}\right) ^{2}e^{i\eta _{+}\theta }-\left( 1+\eta _{+}\right)
^{2}e^{-i\eta _{+}\theta }}\left[ \frac{\left( 1-\eta _{+}\right) ^{2}}{%
1+\eta _{+}}e^{i\eta _{+}\theta }+\frac{\left( 1+\eta _{+}\right) ^{2}}{%
\left( 1-\eta _{+}\right) }e^{-i\eta _{+}\theta }+2\frac{3\eta _{+}^{2}+1}{%
\eta _{+}^{2}-1}e^{-i\theta }\right] ,  \label{transmitted_amplitude_+_z}
\end{equation}%
for the mode that couples to the plus-polarization. For the mode that
couples to the cross-polarization we similarly obtain
\begin{equation}
f_{2}^{R}=\frac{\tilde{h}_{\times }B_{0}}{2}\left( \eta _{\times
}^{2}+1\right) \frac{e^{-i\theta }\left[ e^{i\theta \eta _{\times
}}-e^{-i\theta \eta _{\times }}\right] }{\left( \eta _{\times }-1\right)
^{2}e^{i\theta \eta _{\times }}-\left( \eta _{\times }+1\right)
^{2}e^{-i\theta \eta _{\times }}},  \label{eq-rrr}
\end{equation}%
and
\begin{equation}
f_{2}^{T}=\frac{\tilde{h}_{\times }B_{0}}{2}\left( \frac{\eta _{\times
}^{2}+1}{\eta _{\times }^{2}-1}\right) \frac{\left( \eta _{\times }-1\right)
^{2}e^{i\theta \eta _{\times }}-\left( \eta _{\times }+1\right)
^{2}e^{-i\theta \eta _{\times }}+4\eta _{\times }e^{-i\theta }}{\left( \eta
_{\times }+1\right) ^{2}e^{-i\theta \eta _{\times }}-\left( \eta _{\times
}-1\right) ^{2}e^{i\theta \eta _{\times }}}.  \label{eq-bbb}
\end{equation}%
Here we have introduced the notation $\eta _{+,\times }\equiv
k_{E}^{+,\times }/k$ \ and $\theta =kL$. An example of the magnetic profile
(containing both the transmitted and reflected wave) is given in Fig.\ref{wave_profile} for $kL=40$ and $cB_{0}/E_{\mathrm{cr}}=100$.
%
%
The expressions (\ref{reflected_amplitude_+_z})-(\ref{eq-bbb}) contains all information about the
energy conversion to the different EM-modes. However, to appreciate these
results and the effects due to QED, we must first evaluate some results for
the low-field limit when $\eta _{+,\times }\rightarrow 1$. The squared
coefficient
$\vert f_{1}^{T}\vert ^{2}$,
 proportional to the
energy density of the transmitted wave excited by the $+$-polarization, then
becomes%
\begin{equation}
\left\vert f_{1}^{T}\right\vert ^{2}=\frac{1}{4}\left\vert \tilde{h}%
_{+}\right\vert ^{2}B_{0}^{2}k^{2}L^{2} , 
\end{equation}%
and similarly for the mode excited by the opposite polarization,
\begin{equation}
\left\vert f_{2}^{T}\right\vert ^{2}=\frac{1}{4}\left\vert \tilde{h}_{\times
}\right\vert ^{2}B_{0}^{2}k^{2}L^{2} .
\end{equation}%
Thus we see that the transmitted energy density is directly proportional to
the background energy density. However, this behavior is dramatically
changed when QED-effects are taken into account. The main reason is that the
EM wave dispersion relation is changed in the interaction region (that makes
$\eta _{+,\times }$ deviate from unity) which in turn detunes the excited
wave with the GW. The consequence for the transmitted wave excited by the $%
\times $-polarization is depicted in Fig. \ref{Twavecross}, for $kL=20$ and $kL=100$. The steady increase in the absence of QED is replaced by an oscillatory
behavior, mainly due to the detuning of the GW and EM wave dispersion
relation. Note that we here have normalized the transmission coefficient
with
$|\tilde{h}_{+,\times }|^{2}B_{0}^{2}k^{2}L^{2}$,
such that the coefficient without QED-effects is represented by a straight
line. For a longer interaction region, a smaller mismatch of dispersion
relations are needed for the phase difference to accumulate, and hence the
curve with the lower value of $kL$ ($kL=20$) needs a much higher field
strength before significant QED-effects are seen. A similar point is
illustrated by Fig. \ref{Twaveplus} that depicts the energy density for the co-propagating mode excited by the $+$-polarization. Note that the energy conversion to
this EM-mode is much less affected by the QED effects. The reason is that
the QED-modification of the EM dispersion relation effectively saturates at
a value $cB_{0}/E_{\mathrm{cr}}\sim 10$. Accordingly we have chosen higher
values of $kL$, namely $kL=2000$ and $kL=20000$, which is needed in order to
see the deviation from the classical behavior induced by QED. In addition to
the co-propagating EM modes there are also counter-propagating EM-waves. From a
practical point of view, these are much less significant, since the
counter-propagating modes are always non-resonant with the source GW, and hence
the energy density of these modes does not systematically increase with a
larger interaction region, i.e. increasing $kL$. From a more theoretical
point of view, an interesting effect can be seen in the coefficients (\ref%
{reflected_amplitude_+_z}) and (\ref{eq-rrr}), however. Without QED-effects,
the $+$-polarization does not cause a back-scattered wave, independent of
the value of $kL$, as seen by (\ref{reflected_amplitude_+_z}) when letting $%
\eta _{+}\rightarrow 1$. However, the situation for the $\times $%
-polarization is different, as we find a finite but small counter-propagating mode
from (\ref{eq-rrr}) also in the limit $\eta _{\times }\rightarrow 1$.
%
%

\section{Summary and Conclusion}

In this paper we have studied the interaction between GW:s and EM-waves in
the presence of a strong static magnetic field $B_{0}$, using the
Heisenberg-Euler lagrangian in order to take QED vacuum polarization and
magnetization into account. The high-frequency approximation has been
applied to zeroth order, i.e. all effects of the background curvature has
been neglected, which is permissible 
if the spatial extension of the
interaction region is much smaller than the radius of curvature. The
specific boundary conditions considered is an incoming GW incident on a
static magnetic field with a given extent $L$, which give raise to an
excited EM-wave in the same direction as the GW, as well as one propagating
in the opposite direction. The role of the QED effects is twofold: Firstly,
the coupling strength between the GW:s and the electromagnetic waves are
modified (as described by the coefficients of the right hand side in Eq. (%
\ref{wave_equation_final})). Secondly, the change in phase velocity ($<c$)
of the EM-waves induced by the vacuum polarization, as described by the
expressions ${k_{E}^{\times }}$ and ${k_{E}^{+}}$, destroys the perfect
resonance with the gravitational source wave, which gives a saturation of
the possible energy conservation at a finite value of $L$. These effects are
similar in principle for the $h_{\times }$- and $h_{+}$-polarizations (which
couples to different EM-polarizations), and the dimensionless parameter $%
(cB_{0}/E_{\mathrm{cr}})^{2}kL$ need to reach $(cB_{0}/E_{\mathrm{cr}%
})^{2}kL\sim 10^{5}$ in order for QED effects to be important in both cases.
However, since the QED-modification of the EM-mode excited by the $h_{+}$%
-polarization saturates at a value $cB_{0}/E_{\mathrm{cr}}\sim 10$, a much
higher value of $kL$ is needed for the QED-effects to be significant in this case.

The problem considered here has been highly idealized and has mainly been
motivated by a theoretical interest to study GW and EM-wave interaction in a
strong field environment, allowing for field strengths larger than the
Schwinger critical field $E_{\mathrm{cr}}$. However, we would like to point
out that there is a certain astrophysical relevance of the problem, as the
effect of QED-detuning is found to be of significance for field strengths $%
B_{0}\simeq 3E_{\mathrm{cr}}/c\approx 10^{10}\mathrm{T}$ (see e.g. Fig \ref{Twaveplus}), a
value that has been observed at magnetar surfaces \cite{Magnetar}, although a high GW frequency would be required.

\section{Acknowledgement}
D. Papadopoulos is grateful to DAAD and Aristotle University of Thessaloniki, Greece for their financial support of the research reported here. D. Papadopoulos would also like to thank the staff of the Department of Physics at Ume\aa \ University, Sweden, and Professor K. D. Kokkotas, head of the Department of Theoretical Astrophysics in T\"ubingen, Germany, for the warm hospitality during his stay there, where part of this research was carried out. 

M. Forsberg would also like to thank Professor K. D. Kokkotas for the warm hospitality during his stay in T\"ubingen, where parts of this work was carried out. Furthermore, the authors are greatly indebted to J. Lundin for helpful discussions.
%
%
%
%
%

\appendix

\section{Strong field vacuum polarization and magnetization parameters}
With only a strong magnetic field present
the quantities $\gamma_{{\cal F}}, \gamma_{{\cal G}}, \gamma_{{\cal F}{\cal F}}, \gamma_{{\cal G}{\cal G}}$ and $\gamma_{{\cal F}{\cal G}}$ can be determined analytically, see Ref. \cite{Lundin-2009}. The resulting expressions for these QED-parameters are
\begin{eqnarray}
		\gamma_{{\cal G}} &=& 0, \nonumber \\
		\gamma_{{\cal F}{\cal G}} &=& 0,\nonumber\\
		\gamma_{{\cal F}} &=&
		- \frac{1}{\mu_0} - \frac{\alpha}{2\pi \mu_0}\bras{\frac{1}{3}
		+ 2h^2-8\zeta'(-1,h) + 4h\ln{(\Gamma(h))} - 2h\ln{h}
		+ \frac{2}{3}\ln{h}-2h\ln{2\pi} },\nonumber\\
		\gamma_{{\cal F}{\cal F}} &=&
		\frac{\alpha}{2\pi \mu_0 B^2} \bras{\frac{2}{3}	+ 4 h^2\psi( 1 + h) - 2h -4h^2
		- 4h\ln{\Gamma(h)} + 2h\ln{2\pi}-2h\ln{h} },	\nonumber\\
		\gamma_{{\cal G}{\cal G}} &=& \frac{\alpha}{2\pi \mu_0 B^2}
		\bras{ -\frac{1}{3}- \frac{2}{3}\psi(1+h) - 2h^2 + \frac{1}{3h}+8\zeta'(-1,h)
					- 4h\ln{\Gamma(h)}+2h\ln{2\pi}+2h\ln{h} }, \label{a2}
\end{eqnarray}
where $\alpha	= \frac{e^2}{4\pi \ep_0 \hbar c}$ 
is the fine structure constant, $h = \frac{E_{cr}}{2 c B}$,
$\Gamma(h)$ the gamma function, $\psi(h)$ the digamma function
and $\zeta'(-1,h)$ the first derivative of the Hurwitz zeta function with respect to its first argument.

Furthermore, in the absence of a strong electric field we can calculate the integral in the Lagrangian (\ref{Lagrangian}) analytically. Since there is only a strong magnetic field present we have $b=0$. Thus, to compute the integral in Eq. (\ref{Lagrangian}), we expand the integrand and take the limit as $b\rightarrow 0$, thereby obtaining
\begin{equation}
			I = \int_{0}^{i\infty} \frac{d s}{s^3}e^{-e E_cr s / c }
			\times \bras{ (e a s) \coth{(e a s)}-\frac{(e a s)^2}{3}-1 } .
			\label{integral_lagrangian_1}
\end{equation}
%
%
By changing the variables such that $e a s = z$, dividing the integral into three parts, altering the integration path and using the regulator $z^{\epsilon}$ we obtain
\begin{eqnarray}
	I = 
		\bra{e a}^2 \brac{\int_0^{\infty}dz \ e^{- E_{cr} z / ca} z^{\epsilon-2} \coth{(z)}
		- \int_0^{\infty}dz \ e^{-E_{cr} z / ca }\frac{z^{\epsilon-1}}{3}
		-	\int_0^{\infty}dz \ e^{-E_{cr} z/ ca} z^{\epsilon-3} }
		\label{integral_lagrangian_3}
\end{eqnarray}
Since $E_{cr}/ca = 2h$ we find the first, second and third part of the integral to be
\begin{eqnarray}
		I_1 \equiv \int_0^{\infty}dz \ e^{-2 h z} z^{\epsilon-2} \coth{(z)}
		= \frac{1}{\epsilon}(2h^2+\frac{1}{3}) + (1-C-\ln{2})(2h^2+\frac{1}{3})
		- 4\zeta'(-1,h)-2h\ln{(h)}, \label{int1}
\end{eqnarray}
\begin{equation}
		I_2 \equiv\int_0^{\infty}dz \ e^{-2 h z }\frac{z^{\epsilon-1}}{3}
		= -\frac{1}{3\epsilon}+\frac{1}{3}C+\frac{\ln{(2h)}}{3} ,  \label{int2}
\end{equation}
and
\begin{equation}
		I_3 \equiv\int_0^{\infty}dz \ e^{-2 h z} z^{\epsilon-3}
		= -[\frac{2h^2}{\epsilon}+h^2-2h^2\ln{h}+(1-C-\ln{2})2h^2], \label{int3}
\end{equation}
respectively, where $C$ is Eulers constant.
With Eqs.(\ref{int1}),(\ref{int2}) and (\ref{int3}) we can now rewrite Eq. \reff{integral_lagrangian_1} as
\begin{equation}
			I = (e a)^2 \brac{\frac{1}{3}[1-\ln{2}-\ln{(2h)}]+h^2[2\ln{h}-1]
			- 2h\ln{h}-4\zeta'(-1,h) },  \label{integral_lagrangian_full}
\end{equation}
and thus the Lagrangian (\ref{Lagrangian}) becomes
\begin{eqnarray}
		\mathcal{L} = - \frac{1}{\mu_0}\mathcal{F} - \frac{\alpha B^2}{2\pi \mu_0}
		\brac{ \frac{1}{3}[1-\ln{2}-\ln{(2h)}]+h^2[2\ln{h}-1] - 2h\ln{h}-4\zeta'(-1,h) }
			- A_{\alpha} j^{\alpha}.  \label{Lagrangian_app_A}
\end{eqnarray}
%
Since we have only a magnetic field, ${\cal G}=0$ holds, and the energy-momentum tensor associated with the Lagrangian \reff{Lagrangian_app_A} becomes
\begin{equation}
		T_{\mu\nu} = -\gamma_{{\cal F}}F_{\mu}^{\alpha}F_{\alpha\nu}
				- {\cal L}g_{\mu\nu}. \label{em-tensor_appA}
\end{equation}
Next we proceed by expanding the energy-momentum tensor \reff{em-tensor_appA}. The first order contribution becomes
\be
		\delta T_{\mu\nu} =
			\delta \gamma_{{\cal F}} F_{\mu}^{\alpha}F_{\alpha\nu}
				- \gamma_{{\cal F}}[\delta F_{\mu}^{\alpha}F_{\alpha\nu}
				+ F_{\mu}^{\alpha}\delta F_{\alpha\nu}] - \delta {\cal L}g_{\mu\nu},
				\label{em-tensor_1st_order}
\ee
%
where
\be
  \dl \gamma_{{\cal F}}=\frac{\alpha}{2\pi \mu_0} \bras{ 4 h_0 + 4 \ln \Gamma \bra{h_0}
  + 2 \ln h_0 - 2 \ln 2\pi - 2 - \frac{2}{3 h_0} - 4 h_0 \Psi(h_0) }
  \bra{-h_0 \frac{\dl B_1}{B_0}}, \nonumber
\ee
and $h_0 = E_{cr} / 2 c B_0$,
so the relevant energy-momentum tensor terms in Eq. \reff{gw} becomes
\ber
		\dl T_{11} - \dl T_{22} &=& B_0^2 \dl \gamma_F - 2 \gamma_F B_0 \dl B_1,
				\nonumber \\
		\dl T_{12} &=& \gamma_F B_0 \dl B_2. \label{EM_tensor_variation}
\eer

\section{Ricci-rotation coefficients}
The Ricci-rotation coefficients of a Minkowski spacetime perturbed by a GW propagating in the $\fat e_3$-direction expressed in the tetrad \reff{tetrad} is given by
\begin{eqnarray}
&&\gamma_{11}^0=-\gamma_{22}^0=\gamma_{01}^1=-\gamma_{02}^2=\frac{1}{2c}\dot{h}_{+}, \nonumber\\
&&\gamma_{12}^0=\gamma_{21}^0=\gamma_{02}^1=\gamma_{01}^2=\frac{1}{2c}\dot{h}_{\times}, \nonumber\\
&&\gamma_{31}^1=-\gamma_{32}^2=-\gamma_{11}^3=\gamma_{22}^3=\frac{1}{2}\frac{\partial h_{+}}{\p z}, \nonumber\\
&&\gamma_{32}^1=\gamma_{31}^2=-\gamma_{12}^3=-\gamma_{21}^3=\frac{1}{2}\frac{\partial h_{\times}}{\p z}, \label{ricci}
\end{eqnarray}
to first order in $h_{+,\times}$.
%
%


\begin{thebibliography}{99}
\bibitem{Moortgat2003} J. Moortgat and J. Kuijpers, A\&A \textbf{402}, 905
(2003).

\bibitem{bmd1} M. Marklund, G. Brodin and P. K. S. Dunsby, Astrophys. J.
\textbf{536}, 875 (2000).

\bibitem{Isliker} H. Isliker, I. Sandberg, L. Vlahos, Phys. Rev. D \textbf{74%
}, 104009 (2006).

\bibitem{Papadouplous2001} D. Papadopoulos, N. Stergioulas, L. Vlahos and J.
Kuijpers, A\&A \textbf{377}, 701 (2001).

\bibitem{kallberg2004} A. K\"{a}llberg, G. Brodin and M. Bradley, Phys. Rev.
D \textbf{70}, 044014 (2004).

\bibitem{ignatev} Yu G. Ignat'ev, Phys. Lett. A \textbf{320}, 171 (1997).

\bibitem{brodinmarklund} G. Brodin and M. Marklund, Phys. Rev. Lett. \textbf{%
82}, 3012 (1999).

\bibitem{bmd2} G. Brodin, M. Marklund and P. K. S. Dunsby, Phys. Rev. D
\textbf{62}, 104008 (2000).

\bibitem{BMS2001} G. Brodin, M. Marklund and M. Servin, Phys. Rev. D 63,
124003 (2001).

\bibitem{Balakin2003} A. B. Balakin, V. R. Kurbanova and W. Zimdahl, J.
Math. Phys., \textbf{44}, 5120 (2003)

\bibitem{servinbrodin} M. Servin and G. Brodin, Phys. Rev. D \textbf{68},
044017 (2003).

\bibitem{Servin2000} M. Servin, G. Brodin, M. Bradley and M. Marklund, Phys.
Rev E,\textbf{\ 62}, 8493 (2000).

\bibitem{Papadoupolus2002} D. Papadopoulos, Class Quantum Grav. \textbf{19},
2939 (2002).

\bibitem{MDB2000} M. Marklund, P.K.S. Dunsby, and G. Brodin, Phys. Rev. D
\textbf{62}, 101501 (2000).

\bibitem{Mosquera2002} H. J. M. Cuesta, Phys. Rev. D \textbf{65}, 64009
(2002).

\bibitem{Moortgat2006} J. Moortgat and J. Kuijpers, MNRAS 368 
, 1110 (2006).

\bibitem{Marklund-review} M. Marklund and P. K. Shukla, Rev. Mod. Phys.
\textbf{78}, 591 (2006).

\bibitem{Brodin-pla} G. Brodin, L. Stenflo, D Anderson, M. Lisak, M.
Marklund and P. Johannisson, Phys. Lett A, \textbf{306}, 206 (2003).

\bibitem{Lundin-2009} J. Lundin, Europhys. Lett., \textbf{87,} 31001 (2009).

\bibitem{Valluri} S. R. Valluri, D. R. Lamm and W. J. Mielniczuk, Can. J. Phys., \textbf{71}, 389 (1993).

\bibitem{Magnetar} C. Kouveliotou \textit{et al}., Nature \textbf{393,} 235
(1998).

\bibitem{Schwinger} J. Schwinger, Phys. Rev. \textbf{82}, 664 (1951).

\bibitem{Gies} W. Dittrich and H. Gies, \textit{Probing the Quantum Vacuum} (Springer-Verlag, Berlin) 2000.

\bibitem{EllisElst} G. F. R. Ellis and H. van Elst, \textit{Cosmological models, Theoretical and Observational Cosmology}, ed. M Lachièze-Rey (Dordrecht: Kluwer) (1999).

\bibitem{polarization-comment} It should be noted that although the interaction Eq. (\ref{wave_equation_final}) can
be expressed solely in terms of $\delta B_{1}$ and $\delta B_{2}$, the EM-wave polarization in the presence of strong
QED-effects is nontrivial. For a more complete description, see e.g. Ref. \cite{Lundin-2009}.

\end{thebibliography}
\end{document}